
\magnification=1200
\baselineskip=10pt
\def\lsim{<\kern-2.5ex\lower0.85ex\hbox{$\sim$}\ }
\def\rsim{>\kern-2.5ex\lower0.85ex\hbox{$\sim$}\ }
\overfullrule=0pt
\line{\hfil UR-1274\quad\quad\ \ }
\line{\hfil ER-40425-260\ }
\vskip .50cm
\baselineskip=20pt
\centerline{\bf SELF-DUALITY AND THE KdV HIERARCHY}
\vskip 2cm
\centerline{Ashok Das}
\centerline{and}
\centerline{C. A. P. Galv\~ao \footnote*{Permanent address:
CBPF, Dept. of Fields and Particles, Rua Dr Xavier Sigaud 150,
Urca, 22290 Rio de Janeiro, Brazil}}
\centerline{Department of Physics and Astronomy}
\centerline{University of Rochester}
\centerline{Rochester, NY 14627, USA}
\vskip 2cm
\centerline{\underbar{Abstract}}
\bigskip

We derive the entire KdV hierarchy as well as the recursion
 relations from the self-duality condition on
 gauge fields in four dimensions.
\vfil\eject
\noindent {\bf I. \underbar{Introduction}:}

Most of the integrable equations in 1+1 dimensions
 [1-2] such as the KdV (Kortweg
-deVries) equation, the Sine-Gordon equation, the nonlinear Schr\"odinger
equation, can be derived from SL(2,{\bf R}) valued gauge fields in four
dimensions by requiring their field strengths to be self-dual
 [3-5].  More
recently, even the 2+1 dimensional KP (Kadomtsev-Petviashvili) equation
 [6] has been derived from a self-duality condition imposed on SL(2,{\bf R})
valued gauge fields in six dimensions [7].  The solutions to the self-duality
conditions are extremely restrictive and even though individual integrable
equations (e.g. KdV) can be obtained from these conditions, it has proved
infinitely difficult to obtain from these the hierarchy of equations
associated with a given equation.

In this letter, we show how the entire KdV hierarchy of equations along
with the appropriate recursion relations can be obtained from the
self-duality condition
 on gauge fields in four dimensions.  In sec. II we give a
derivation of the KdV hierarchy based on an Abelian zero curvature
condition.  We would like to emphasize that the standard zero curvature
derivation of the KdV hierarchy [8,2] uses the Lie group SL(2,{\bf R}).  This
derivation based on the Abelian symmetry
 is, therefore, new and is useful in the discussion of the self-
duality condition.  In sec. III we show how the KdV hierarchy can be
obtained from the self-duality condition imposed on Abelian gauge fields in
four dimensions.  In sec. IV we derive the KdV hierarchy from SL(2,{\bf R})
valued gauge fields in four dimensions by imposing the self-duality
condition.  We present our conclusions in sec. V.
\medskip
\noindent {\bf II. \underbar{KdV Hierarchy}:}

The derivation of the KdV hierarchy from a zero curvature condition
associated with the group SL(2,{\bf R}) in 1+1 dimension is quite standard
 [8,2].
 In this section, we show how the hierarchy can be obtained from a
vanishing field strength associated with an Abelian gauge field in 1+1
dimension.  Let $A_\mu (x), \> \mu = 0,1$, denote an Abelian gauge field in
1+1 dimension.  Then, the zero curvature condition reads
$$F_{01} = \partial_0 A_1 - \partial_1 A_0 = 0$$
$${\rm or,}\qquad \partial_0 A_1 = \partial_1 A_0 \eqno(1)$$
where $\partial_0 = {\partial \over \partial t}$ and $\partial_1 =
{\partial \over \partial x} = \partial$.  It is clear that if we identify
$$\eqalign{A_1 (x,t) &= u (x,t)\cr
A_0 (x,t) &= {\delta H_n \over \delta u(x,t)}\cr}\eqno(2)$$
where $u(x,t)$ is the dynamical variable of the KdV hierarchy and
$H_n[u]$ is the nth conserved charge, then the zero curvature condition of
Eq. (1) would take the form
$${\partial u \over \partial t} = {\partial \over \partial x}
\> {\delta H_n \over \delta u(x,t)} \eqno(3)$$
which we recognize to be the nth order equation in the KdV hierarchy.

Let us further note that if we choose
$$\eqalign{A_1 (x,t) &= u(x,t)\cr
A_0 (x,t) &= \big( - 2 \lambda^2 + {1 \over 2}
\> \partial^{-1} \big( \partial^3 + 2 (2 u + u \partial) \big) \big)
C(u)\cr}\eqno(4)$$
where $\lambda$ is an arbitrary constant parameter (commonly known as the
spectral parameter), $\partial^{-1}$ is the formal inverse of the
derivative operator (can be thought of as a space integration) and $C(u)$
is a functional of the dynamical variable $u(x,t)$, then the zero curvature
condition would give
$$\eqalign{\partial_0 A_1 &= \partial_1 A_0 = \partial A_0\cr
{\rm or,}\quad {\partial u \over \partial t} &= - 2 \lambda^2 \partial
 C(u) + {1 \over 2}\> \big( \partial^3 + 2 (\partial u + u \partial )
\big) C(u)\cr}\eqno(5)$$
The derivation of the KdV hierarchy as well as the recursion relations is
now standard (see for example, ref. 2 p137-138).  We note here that it is
 the biHamiltonian structure of the KdV hierarchy which leads
 to the derivation of these equations from two distinct zero curvature
conditions based on the groups U(1) and SL(2,{\bf R}).
\vfil\eject
\medskip
\noindent {\bf III. \underbar{Self-duality condition for U(1) symmetry}:}

Let us next consider gauge fields in a four dimensional space with
signature (2,2).  Thus, we can identify $x^0 = t,\> x^1 = x,\>
 x^2 = \tilde t, \> x^3 = \tilde x$.  If we choose
$$\epsilon^{0123} = 1 \eqno(6)$$
then it follows that in this space
we will have
$$\epsilon_{0123} = 1 \eqno(7)$$
The self-duality condition for the field strengths associated with any
group are given by
$$F_{\mu \nu} = {1 \over 2}\> \epsilon_{\mu \nu}^{\>\> \lambda \rho}
F_{\lambda \rho} \qquad \mu, \nu, \lambda, \rho = 0,1,2,3 \eqno(8)$$
where for matrix gauge potentials,
 the field strength is, in general,   defined to be
$$F_{\mu \nu} = \partial_\mu A_\nu - \partial_\nu A_\mu + [A_\mu,
A_\nu] \eqno(8a)$$

For an Abelian gauge field, the
 self-duality relations can be written out in detail as
$$F_{01} = - F_{23}$$
$${\rm or,}\quad \partial_0 A_1 - \partial_1 A_0 = - \left( \partial_2 A_3
- \partial_3 A_2 \right) \eqno(9)$$
$$F_{02} = - F_{13}$$
$${\rm or,}\quad \partial_0 A_2 - \partial_2 A_0 = - \left( \partial_1
A_3 - \partial_3 A_1 \right) \eqno(10)$$
$$F_{03} = - F_{12}$$
$${\rm or,}\quad \partial_0 A_3 - \partial_3 A_0 = - \left( \partial_1
A_2 - \partial_2 A_1 \right) \eqno(11)$$
If we now assume that all the gauge potentials are independent of the extra
coordinates $(x^2,x^3)$ and identify $A_2 = A_3$, then the self-duality
conditions in Eqs. (9)--(11) reduce to
$$\eqalignno{&\partial_0 A_1 - \partial_1 A_0 = 0 &(12)\cr
&\big( \partial_0 + \partial_1 \big) A_2 = 0 &(13)\cr}$$
The solution to Eq. (13) is quite simple, namely,
$$A_2 (x,t) = A_2(t-x) \eqno(14)$$
It is also clear that with the choice in Eq. (4) for the gauge potentials
 $A_0$ and $A_1$, we obtain the KdV hierarchy from Eq. (12).  This shows
how the entire
 KdV hierarchy can be obtained from the self-duality condition on an
Abelian gauge field in four dimensions.

For completeness and ease of comparison with other references, let us note
that the self-duality condition as given in refs.
 [5] and [7] take the following form
for an Abelian field.
$$\eqalignno{&\partial_t Q - \partial_x H = 0 &(15)\cr
&\partial_x (Q- P ) = 0 &(16)\cr}$$
In this case, if we identify
$$\eqalign{Q &= P = u\cr
H &= \big( - 2 \lambda^2 + {1 \over 2}\> \partial^{-1} \big( \partial^3 + 2
(\partial u + u \partial )\big)\big) C(u)\cr}\eqno(17)$$
then Eq. (16) is automatically satisfied while Eq. (15) leads to the KdV
hierarchy.
\medskip
\noindent {\bf IV. \underbar{Self-duality condition for SL(2,R) symmetry}:}

Let us consider again the four dimensional space with (2,2) signature
 but gauge fields belonging
 to the group SL(2,{\bf R}).  In this case, the self-duality conditions in
Eq. (8) can be written out explicitly as
$$F_{01} = - F_{23}$$
$${\rm or,}\quad \partial_0 A_1 - \partial_1 A_0 +
[A_0, A_1] = - \left( \partial_2
 A_3 - \partial_3 A_2 + [A_2, A_3]\right) \eqno(18)$$
$$F_{02} = - F_{13}$$
$${\rm or,}\quad \partial_0 A_2 - \partial_2 A_0 + [A_0, A_2] = - \left(
\partial_1 A_3 - \partial_3 A_1 + [A_1, A_3 ]\right) \eqno(19)$$
$$F_{03} = - F_{12}$$
$${\rm or,}\quad \partial_0 A_3 - \partial_3 A_0 + [A_0, A_3] = -
\left( \partial_1 A_3 - \partial_2 A_1 + [A_1, A_2]\right) \eqno(20)$$
Here we assume the gauge potentials as belonging to the $2 \times 2$ matrix
representation of SL(2,{\bf R}).

Once again, let us assume that all the gauge potentials are independent of
the additional coordinates $(x^2, x^3)$.  Furthermore, if we identify
$$A_2 (x,t) = A_3 (x,t) \eqno(21)$$
the the self-duality conditions of Eqs. (18)--(20) reduce to
$$\eqalignno{&\partial_0 A_1 - \partial_1 A_0 + [A_0, A_1] = 0 &(22)\cr
&\big( \partial_0 + \partial_1 \big) A_2 + [A_0 + A_1, A_2 ] = 0 &(23)\cr}
$$
If we now choose the SL(2,{\bf R}) potentials as
$$\eqalign{A_1 &= \pmatrix{\lambda &-u\cr
\noalign{\vskip 6pt}%
1 &-\lambda\cr}\cr
\noalign{\vskip 6pt}%
A_0 &= \pmatrix{\lambda C(u) - {1 \over 2}\> \partial C(u)
&- {1 \over 2}\> \partial^2 C(u) + \lambda \partial C(u) - u C(u)\cr
\noalign{\vskip 6pt}%
C(u) & - \lambda C(u) + {1 \over 2}\> \partial C(u)\cr}\cr}\eqno(24)$$
then Eq. (22) will give the equation generating the KdV hierarchy, namely,
$${\partial u \over \partial t} = - 2 \lambda^2 \partial C(u) +
{1 \over 2}\> \left( \partial^3 + 2
(\partial u + u \partial)\right) C(u) \eqno(25)$$
It is also easy to solve Eq. (23) at least formally as
$$A_2 (x^+,x^-) = A_3 (x^+,x^-) =
e^{- \int^{x^+}dx^{\prime +}A_+ (x^{\prime +},x^-)}
A_2 (0,x^-) e^{\int^{x^+} dx^{\prime \prime +}A_+ (x^{\prime \prime +} ,
 x^-)}\eqno(26)$$
where we have defined
$$\eqalign{x^\pm &= t \pm x\cr
A_+ &= A_0 + A_1\cr} \eqno(27)$$
Note that the solution in Eq. (26) holds for any nontrivial matrix function
$A_2 (0,x^-)$.  In particular, we may choose it to be a constant matrix
belonging to SL(2,{\bf R}).  It is clear that any choice for
$A_2 (0,x^-)$ does not effect the dynamical equation in Eq. (25).  This
shows, therefore, that the KdV hierarchy as well as the recursion relations
can also be obtained from the self-duality condition imposed on SL(2,{\bf R})
valued gauge fields in four dimensions with (2,2) signature.
\medskip

\noindent {\bf V. \underbar{Conclusion}:}

We have given a derivation of the KdV hierarchy based on the zero curvature
condition associated with a U(1) symmetry group.  We have also  obtained
the KdV hierarchy from a self-duality condition on gauge fields in four
dimensions (with (2,2) signature) belonging to U(1) as well as to SL(2,
{\bf R}) symmetry groups.  The extension of our results to other 1+1
dimensional integrable systems should be quite straightforward.

This work was supported in part by the U. S. Department of Energy Grant No.

\noindent DE-FG02-91ER40685 and by CNPq, Brazil.
\vfil\eject
\noindent {\bf \underbar{References}:}
\item{1.} L. D. Faddeev and L. A. Takhtajan, \lq\lq Hamiltonian methods
 in the theory of solitons", Springer (Berlin 1987).
\item{2.} A. Das, \lq\lq Integrable Models", World Scientific (Singapore
 1989).
\item{3.} R. S. Ward, Nucl. Phys. {\bf B236} (1984) 381;
 Philos. Trans. R. Soc.
{\bf A315} (1985) 451; also in \lq\lq Field theory, quantum gravity and
strings", vol. 2, eds. H. J. deVega and N. Sanchez, p106.
\item{4.} L. J. Mason and G. A. J. Sparling, Phys. Lett. {\bf A137} (1989)
29.
\item{5.} I. Bakas and D. A. Depireux, Intern. J. Mod. Phys. {\bf A7}
(1992) 1767.
\item{6.} B. B. Kadomtsev and V. I. Petviashvili, Sov. Phys. Dokl. {\bf 15}
 (1971) 539.
\item{7.} A. Das, Z. Khviengia and E. Sezgin, Phys. Lett. {\bf B289} (1992)
347.
\item{8.} S. S. Chern and C-K. Peng, Manuscripta Mathematica {\bf 28}
(1979) 207.
\bye